\documentclass[]{spie}  

\usepackage{float}
\usepackage{amsmath,amsfonts,amssymb}
\usepackage{graphicx}
\usepackage[colorlinks=true, allcolors=blue]{hyperref}
\DeclareUnicodeCharacter{2212}{-}

\title{APLC-Optimization: an apodized pupil Lyot coronagraph design survey toolkit}

\author[a]{Bryony F. Nickson}
\author[a]{Emiel H. Por}
\author[a]{Meiji M. Nguyen}
\author[a]{R{\'{e}}mi Soummer}
\author[b]{Iva Laginja}
\author[a]{Ananya Sahoo}
\author[a]{Laurent Pueyo}
\author[a]{Kathryn St.Laurent}
\author[c]{Mamadou N’Diaye}
\author[d]{Neil T. Zimmerman}
\author[a]{James Noss}
\author[a]{Marshall Perrin}

\affil[a]{Space Telescope Science Institute, 3700 San Martin Dr, Baltimore, United States}
\affil[b]{LESIA, Observatoire de Paris, Universit\'{e} PSL, Sorbonne Universit\'{e}, Universit\'{e} Paris Cit\'{e}, CNRS, 5 place Jules Janssen, 92195 Meudon, France}
\affil[c]{Observatoire de la Côte d'Azur, Nice, France}
\affil[d]{NASA Goddard Space Flight Center, Goddard, United States}

\authorinfo{Further author information: Send correspondence to Bryony Nickson: E-mail: bnickson@stsci.edu, Telephone: +1 410 338 6739}

\begin{document} 
\maketitle

\begin{abstract}
We present a publicly available software package developed for exploring apodized pupil Lyot coronagraph (APLC) solutions for various telescope architectures. In particular, the package optimizes the apodizer component of the APLC for a given focal-plane mask and Lyot stop geometry to meet a set of constraints (contrast, bandwidth etc.) on the coronagraph intensity in a given focal-plane region (i.e. dark zone). The package combines a high-contrast imaging simulation package (HCIPy\cite{hcipy}) with a third-party mathematical optimizer (Gurobi) to compute the linearly optimized binary mask that maximizes transmission. We provide examples of the application of this toolkit to several different telescope geometries, including the Gemini Planet Imager (GPI) and the High-contrast imager for Complex Aperture Telescopes (HiCAT) testbed. Finally, we summarize the results of a preliminary design survey for the case of a $\sim$6~m aperture off-axis space telescope, as recommended by the 2020 NASA Decadal Survey, exploring APLC solutions for different segment sizes. We then use the Pair-based Analytical model for Segmented Telescope Imaging from Space (PASTIS) to perform a segmented wavefront error tolerancing analysis on these solutions.

\end{abstract}

\keywords{coronagraph design, apodized pupil Lyot coronagraph, high-contrast imaging, LUVex, GOMaP, segmented telescopes}

\section{INTRODUCTION}
\label{sec:intro}  

Direct imaging of exoplanets, including those which might be habitable, is a major goal in astronomy for the next decade. Indeed, this is one of the key recommendations of the 2018 National Academies of Sciences’ Exoplanet Science Strategy Report\cite{nas} and of the recent Astro2020 decadal survey: Pathways to Discovery in Astronomy and Astrophysics for the 2020s\cite{astro2020}, which placed a ``IR/O/UV telescope optimized for observing habitable exoplanets and general astrophysics'' as their highest priority recommendation in the frontier category for space.



The complexity of directly imaging exoplanets stems from two main challenges: the need to resolve substellar companions at very small angular separations from their host stars, as well as the large flux ratio between the planet and star. For these reasons, we require large-aperture telescopes equipped with coronagraphic instruments capable of extreme starlight suppression. In this study we focus on a well tested coronagraph type that is currently the basis for several ground-based high-contrast imaging instruments, the Apodized Pupil Lyot Coronagraph (APLC) - a Lyot-style coronagraph that suppresses starlight through a series of amplitude operations on the on-axis electric field.

\begin{figure}[h]
    \centering
    \includegraphics[width=0.9\linewidth]{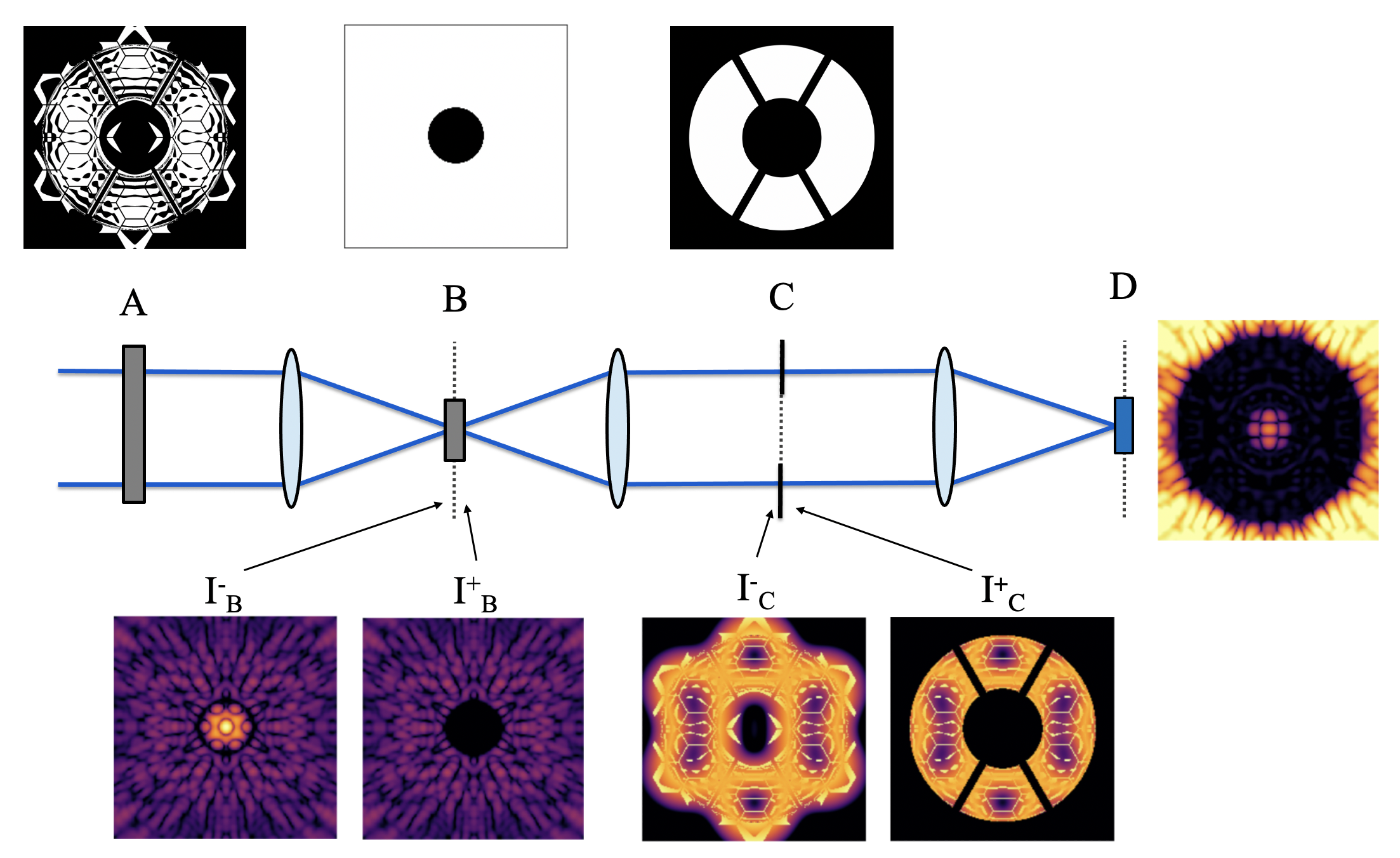}
    \caption{Optical Layout of the APLC. The blue rays describe the path of the on-axis starlight through the system. The binary masks at the top show the segmented pupil with the apodizer, FPM and LS that constitute the APLC. The image pairs at the bottom show the light in each of the coronagraphic planes: before and after the FPM (plane B), and before and after the LS (plane C, normalized to the peak intensity of the unocculted PSF). The image to the far right shows the normalized irradiance of the post-coronagraphic stellar PSF at the detector. All images are on the same logarithmic scale and only show the on-axis light.}
    \label{fig:schematic}
\end{figure}

A schematic diagram of the APLC is shown in Figure~\ref{fig:schematic}, with an example of on-axis light propagating through each of the coronagraphic masks. The APLC optical layout consists of an apodizer in the pupil plane A that modulates the optical beam in amplitude, a downstream focal plane mask (FPM) in plane B that occults the on-axis point-spread-function (PSF) core, and a Lyot stop (LS) in the relayed pupil plane C that blocks the light diffracted by the FPM to form the coronagraphic image on a detector located in the re-imaged focal plane D.

In this work, we present a publicly available software toolkit developed to explore the design of APLCs. We provide a broad overview of the APLC-optimization package and the approaches used in Section~\ref{sec:aplc_opt}. In Section~\ref{sec:examples}, we provide examples of existing coronagraph designs optimized by our software. In Section~\ref{sec:survey}, we present an application of the toolkit for a parameter study in the case of a $\sim$6m aperture off-axis space telescope, producing APLC solutions for different segment sizes and analyzing their performance (contrast, inner and outer working angles, throughput), as well as WFE tolerances in Section~\ref{sec:pastis}. We conclude with Section~\ref{sec:conclusion}.

\section{The APLC-Optimization toolkit}
\label{sec:aplc_opt} 
APLC-Optimization is a software toolkit developed by the SCDA (Segmented Coronagraph Design $\&$ Analysis) research team at the Space Telescope Science Institute (STScI) for the purpose of exploring APLC solutions for arbitrary telescope apertures. It uses a numerical optimization method to find transmission-maximizing binary apodizers for a given combination of design constraints (such as the contrast goal, dark zone inner working angle and outer working angle, spectral bandwidth, telescope pupil, occulting mask, and Lyot stop profile). The object-oriented approach of the APLC-optimization toolkit simplifies the interface for extensive parameter space studies and enables flexibility for implementing various mask architectures, including features such as mirror segment gaps and struts. 

The simulation of all propagations is performed with HCIPy\cite{hcipy}, using the semi-analytical Lyot coronagraph propagation method described in Soummer et al. (2007)\cite{2007OExpr..1515935S}. For efficiency, the code only propagates light to the part of the dark zone that is used for the constraints by treating the area outside of this focal-plane mask as fully transmissive. For the numerical optimization itself, we rely on the Gurobi\footnote{\url{https://www.gurobi.com/products/gurobi-optimizer/}} solver to compute the apodizer mask with maximized off-axis transmission. 

The APLC-Optimization package includes complete documentation of all classes and functions. The core software package, example notebooks, and documentation are all publicly hosted through the STScI GitHub organization\footnote{\url{https://github.com/spacetelescope/aplc_optimization}}.

\subsection{Apodizer Optimization Method}
\label{sec:ap_opt}
A detailed description of our complete numerical optimization method is provided in Por et al. 2020\cite{por2020aplc}. Specifically, the optimization problem maximizes the peak of the non-coronagraphic image while simultaneously constraining the intensity of the broadband coronagraphic image in the dark-zone region. It forces it to be lower than the desired raw contrast limit, which is measured as the relative intensity with respect to the peak intensity of the non-coronagraphic image. 

Por et al. 2020 further describes the methods we use to reduce the dimensionality of the optimization problem and significantly speed up the optimization process. These include exploiting symmetries in the underlying problem and a progressive refinement algorithm that relies on the binary structure of the masks to progressively upscale low-resolution solutions, at minimal cost. Altogether, these techniques significantly reduce computation time and memory usage, increasing the ease of larger and more intricate mask optimizations and parameter space explorations\cite{por2020aplc}. 


\subsection{Parameter space surveys}
The toolkit features a survey mode designed to simplify the organization, execution, and evaluation of extensive parameter space studies. Design surveys are performed using ``optimization launchers'', wherein the user defines a set of design parameters to be surveyed. These include a list of telescope aperture specifications, Lyot stop dimensions, focal-plane mask sizes, tolerance constraints, and contrast and bandwidth goals. At the same time, the user can also set the parameters for the optimization method, such as whether to turn on the adaptive algorithm, ignore certain symmetries in the optimization problem, or set the number of threads to be used by the solver. Any unspecified parameters are set to reasonable default values. 

Launcher templates are provided for a number of realistic telescope configurations, such as LUVOIR-A\cite{luvoir}, HiCAT\cite{2022SPIEremi}, and Gemini/GPI\cite{2022SPIEmeiji}. Once launched, the toolkit produces all the necessary input files defining the telescope apertures and intermediate masks; it then calls a routine to write a collection of ``driver'' files (one for each parameter combination) to be executed in queued batches on a computing server. These driver files, along with all logging files, are bundled together in a ``survey'' folder, providing a complete paper trail for the entire collection of optimizations so that each distinct optimization can be checked or repeated in the future. Once each optimization program completes, the apodizer solution is saved to disk, and an analysis module is automatically run. This analysis module produces a PDF file containing a design summary and several relevant plots so that the design can be evaluated immediately after optimization.


\subsection{Design robustness to Lyot stop misalignments}
\begin{figure} [t]
   \begin{center}
   \begin{tabular}{c} 
   \includegraphics[width=0.9\linewidth]{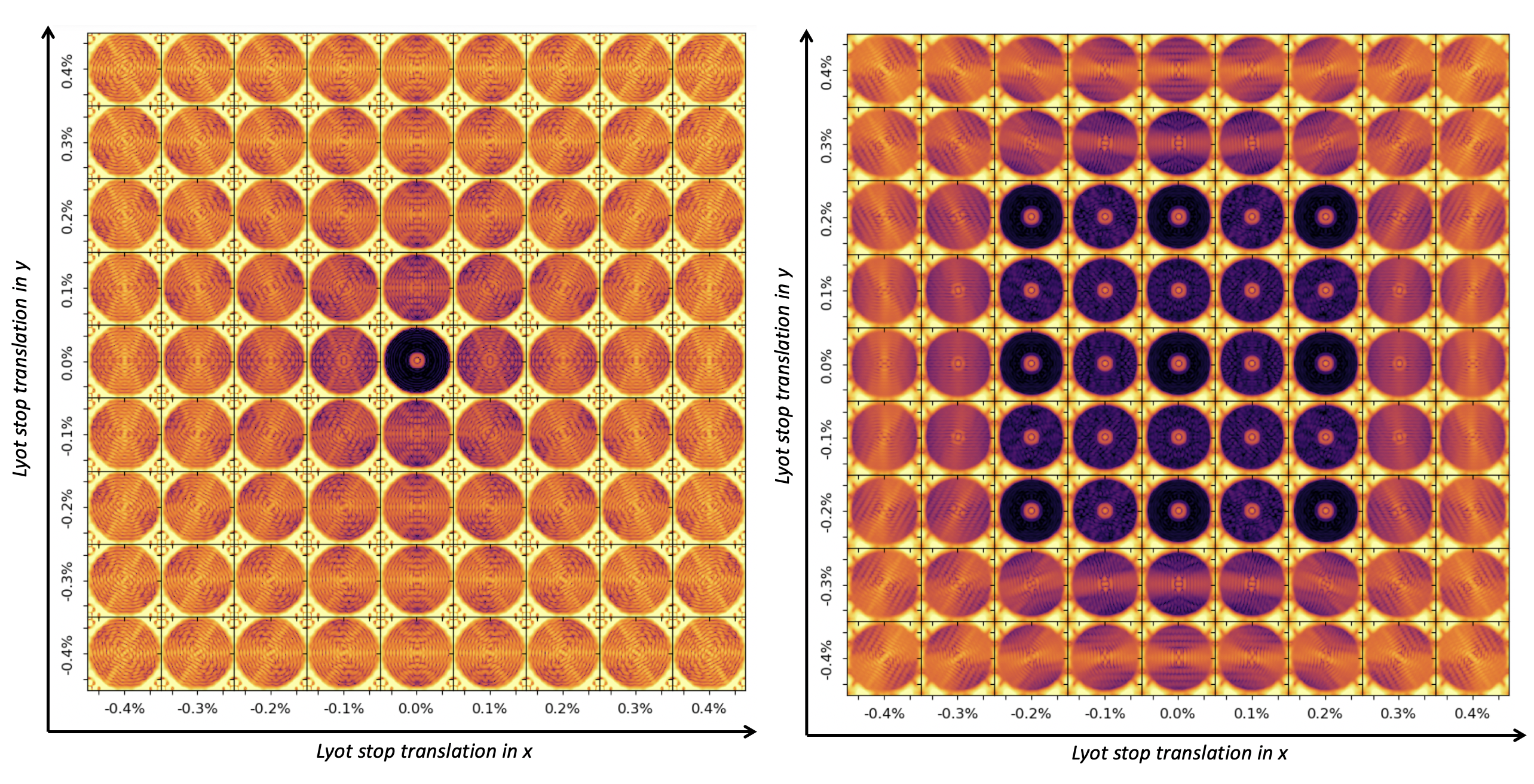}
   \end{tabular}
   \end{center}
   \caption[example] 
   { \label{fig:lsroh} Lyot stop misalignment sensitivity for two 1-Hex apodizer masks (see Section~\ref{sec:survey}) one without (\textit{left}) and one with (\textit{right}) built-in robustness to Lyot stop translations. The two plots show a grid of post-coronagraphic images for different Lyot stop positions. The colorbar for all images stretches in logarithmic scale from $10^{-4}$ to $10^{-9}$. The axes show the translation of the Lyot stop in the x and y directions. \textit{Left:} Non-robust APLC design, where only the perfectly aligned LS yields a deep dark hole (DH). \textit{Right:} Robust APLC design with lesser sensitivity to LS misalignments. This design yields a deep-contrast DH for a set of misalignment states of the LS.}
   \end{figure}

Future space-based telescopes will require designs that operate in broadband light and are insensitive to fabrication and alignment errors. The toolkit approaches the development of APLC designs robust to Lyot stop misalignments by optimizing the apodizer for multiple translated versions of the Lyot stop simultaneously. Figure \ref{fig:lsroh} shows the Lyot stop misalignment sensitivity for APLC designs without (left) and with (right) built-in Lyot stop robustness properties. For the robust design (right panel in Figure~\ref{fig:lsroh}), the optimization is performed using a set of 9 Lyot stops in a 3 x 3 grid, centred on the nominal Lyot stop position. The two figures show a grid of the coronagraphic images for different translations of the Lyot stop mask. While the non-robust design only produces a dark zone when the Lyot stop is centered on its optimal on-axis position, the design optimized with increased robustness to Lyot stop misalignment produces a dark zone for multiple, translated versions of the Lyot stop. 

\section{Case studies}
\label{sec:examples} 
In this section, we highlight two usage cases of the APLC-Optimization package to produce optical designs for current-generation high-contrast imaging instruments. The first usage case, HiCAT, shows that the toolkit can (1) handle segmented-telescope pupils with non-simple geometries that include features such as support struts (``spiders'') and central obscurations, and (2) produce designs capable of achieving competitive raw contrast levels necessary for future large space-based coronagraphic missions. The second usage case, GPI, shows that the toolkit can perform similar optimizations on ground-based high-contrast imaging instruments.

\subsection{HiCAT}
\label{sec:hicat}

The High-contrast imager for Complex Aperture Telescopes\cite{2022SPIEremi} is a coronagraphic instrument and optical design testbed developed to simulate and study segmented aperture geometries for applications on future large space-based observatories. It features a 37 hexagonal segment IrisAO deformable mirror (DM) which acts as a primary mirror simulator, two Boston Micromachines Kilo-DMs used to perform closed-loop wavefront sensing and control, a Zernike mask and wavefront sensor used to perform low-order aberration wavefront control \cite{2022SPIEraphael}, and modular optics mounts which can accommodate different mask geometries in each of the optical planes of propagation of the instrument - including an apodizer mount, a focal-plane mount (e.g., pinhole, knife-edge), and a LS mount. The APLC-Optimization toolkit is our primary software infrastructure for developing apodizer designs for the HiCAT testbed (see Figure~\ref{fig:case_studies} for an example HiCAT apodizer design). 
We have used the package to produce mask designs which can achieve nominal raw contrasts on the order of $10^{-8}$ to $10^{-9}$ in an annular dark zone down to an IWA of 4 $\lambda /D$ over a spectral bandpass of up to 15\%. HiCAT is an important step on the path to achieving the nominal raw contrast of $10^{-10}$, which is the aspirational target of future space-based coronagraphic instrument concepts \cite{astro2020}.

\subsection{GPI}
\label{sec:GPI}

\begin{figure} [h]
   \begin{center}
   \begin{tabular}{c|c} 
   \hspace{2mm}
   \includegraphics[height=6.6cm]{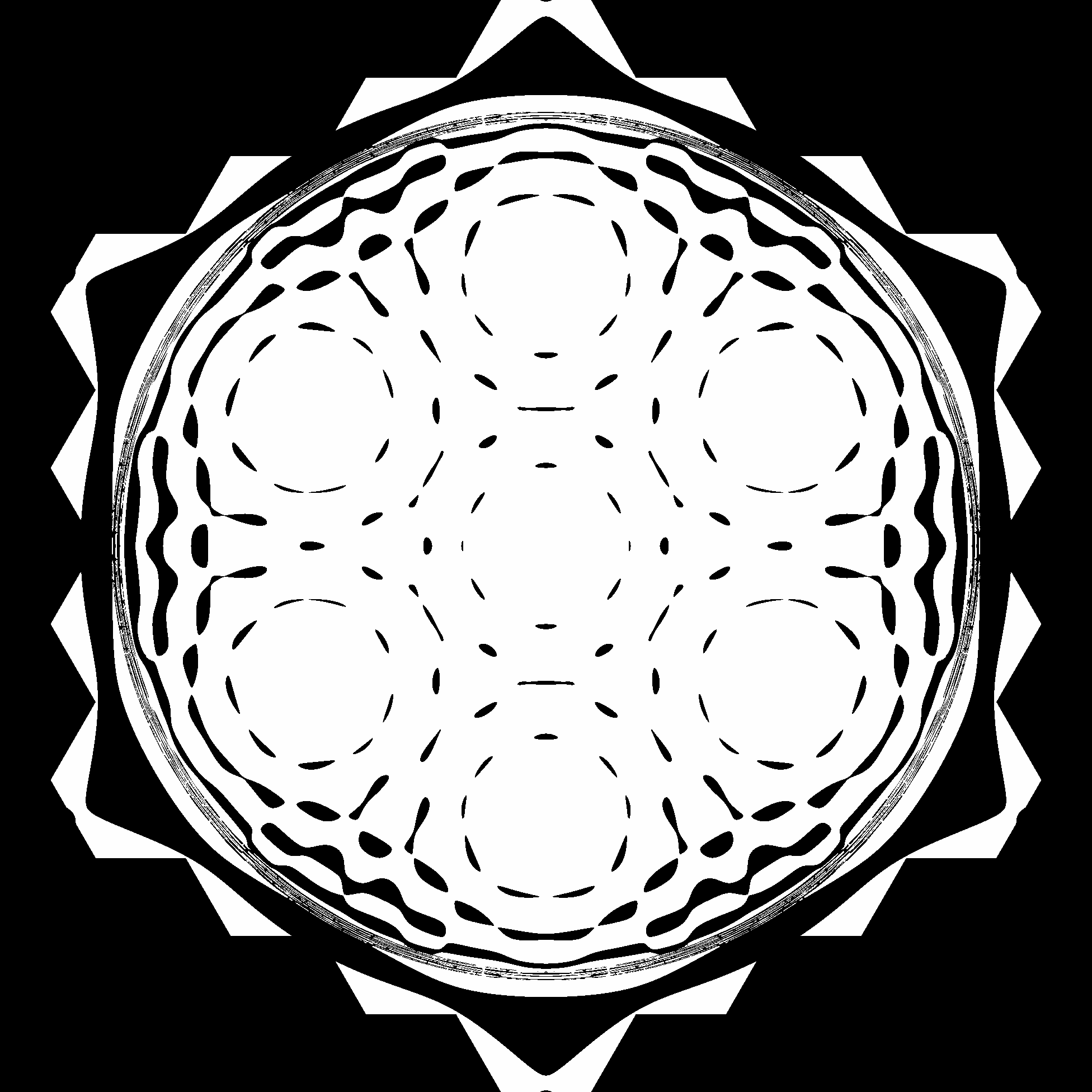} \hspace{1.1cm} & 
   \includegraphics[height=6.8cm]{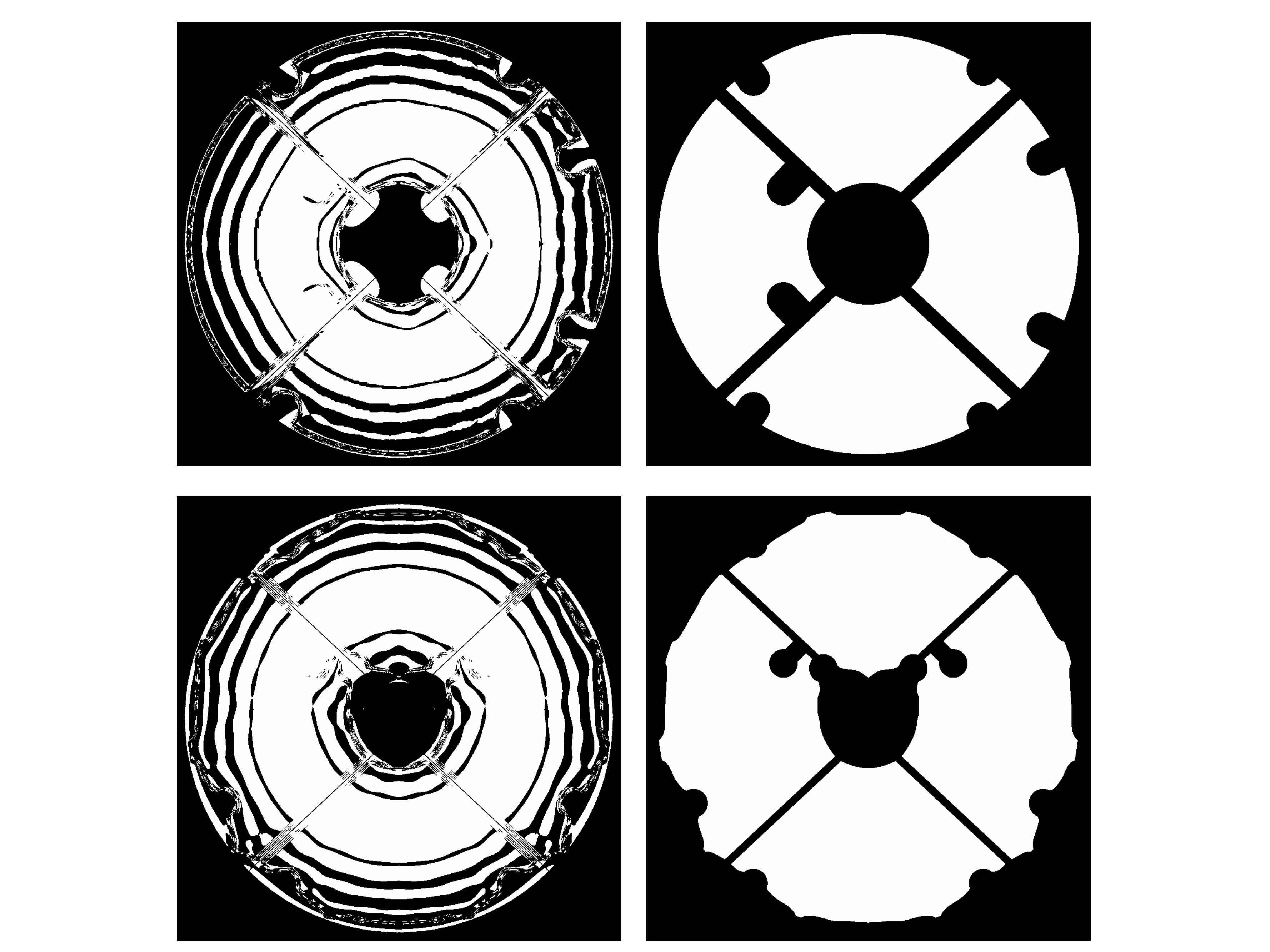}\\
   \end{tabular}
   \end{center}
   \caption[example] 
   { \label{fig:case_studies}\textit{Left:} A binary transmissive apodizer mask developed for HiCAT (optimized at a resolution of 1944 x 1944 pixels and a $\pm$0.3$\%$ robustness to Lyot stop translations). \textit{Right:} Binary transmissive apodizer masks (left column) and Lyot stops (right column) developed for GPI 2.0. \textit{Top row}: One of the new GPI 2.0 apodizer designs, ``LS03Symm'', that is paired with a LS based on a horizontally symmetrized version of the ``080m12 03'' Lyot stop mask previously installed on GPI 1.0. \textit{Bottom row}: Another one of the new GPI 2.0 apodizer designs, ``DualPlaneSymm'', which is paired with a LS based on a symmetrized version of the ``DualPlane'' GPI 2.0 LS. }
   \end{figure}

The Gemini Planet Imager\cite{2014PNAS..11112661M} is an integral field spectrograph and coronagraph that is in the process of being upgraded and moved from its current mount behind the Cassegrain focus of the 8.1m Gemini South Telescope to a similar position on the telescope's twin, Gemini North, as part of the instrument's upgrade process - GPI 2.0 \cite{2020SPIE11447E..1SC}. The APLC-Optimization toolkit was recently used to develop new apodizer and Lyot stop masks as part of the GPI 2.0 upgrade\cite{2022SPIEmeiji, 2022SPIEemiel}. These updated apodizers were optimized to achieve better high-contrast imaging performance, such as deeper raw contrast at the IWA of the dark zone, and increased robustness to Lyot stop misalignments. Some of the matching LS masks developed to go with the new apodizers (see Figure~\ref{fig:case_studies}) were intentionally symmetrized in order to reduce the computational overhead of the optimization (by up to a factor of 4, as described in Section~\ref{sec:ap_opt}). This allowed the optimizations to bypass a memory bottleneck encountered in the optimization process caused by the asymmetry of the original LS designs due to the masking of the dead actuators on the Gemini Telescope's deformable mirror. These new coronagraphic masks have been lithographically printed by Advanced Nanophotonics Inc.\footnote{\url{https://www.advancednanophotonics.com/}}\cite{Hagopian10} and will be ready for use when GPI 2.0 goes on-sky in 2023.

\section{Segment Size Survey for a 6~m off-axis segmented space telescope}
\label{sec:survey} 
Segment size is a key parameter in the design of segmented telescope apertures. Determining the optimal number of segments to fill a particular aperture size involves a complicated trade-off between many factors, including diffraction effects, fabrication and complexity of individual support, versus the complexity of telescope alignment and control. Concerning the future Large IR/Optical/UV telescope recommended by the NASA Astro 2020 Decadal survey, the committee concluded that ``a target off-axis inscribed diameter of approximately 6 meters provides an appropriate balance between scale and feasibility.''\cite{astro2020}. In order to help optimize such a configuration, we conduct a preliminary study of APLC performance for apertures with a wide range of segment sizes, at a fixed telescope inscribed diameter of 6~m.

\begin{figure} [h]
   \begin{center}
   \begin{tabular}{c} 
   \includegraphics[width=\linewidth]{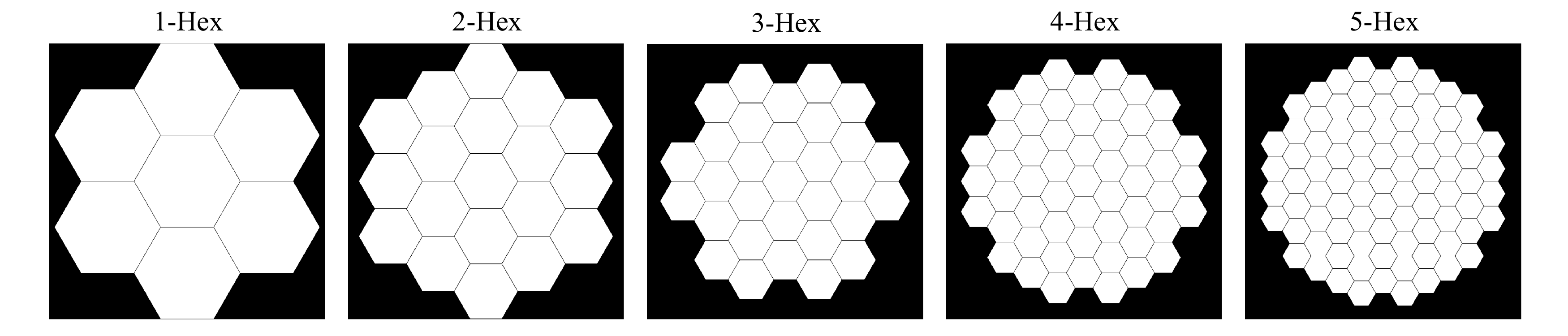}
   \end{tabular}
   \end{center}
   \caption[example] 
   { \label{fig:hex_aps} The five unobscured hexagonal segmented-aperture designs considered in the segment size survey discussed in Section~\ref{sec:survey}. Each aperture consists of N = 1, 2, 3, 4, and 5 rings of hexagonal mirror segments surrounding a central segment. The designs are circularized by exclusion of segments whose centers lie at a distance greater than 85$\%$ of the pupil diameter. All have an inscribed diameter of $\sim$6~m and 60~mm wide segment gaps (note that the segment gaps in these images are padded for display purposes). \textit{From left:} 1-Hex, 2-Hex, 3-Hex, 4-Hex, and 5-Hex.}
   \end{figure}

In this section, we summarize a survey of APLC solutions for five off-axis hexagonal segmented apertures made of segments with varying sizes. Figure \ref{fig:hex_aps} shows the five reference apertures considered, with decreasing segment size from left to right. The apertures are comprised of hexagonal segments organized in N = 1, 2, 3, 4, and 5 hexagonal concentric rings around a central segment. For the N = 3, 4 and 5 ring apertures some of the segments in the outer rings have been omitted in order to maximize the diameter of the inscribed circle with respect to the overall diameter. This results in a circularization of the pupils which has been shown to be advantageous for coronagraphy\cite{St_Laurent_2018}.

To be in line with the recommendations of the Astro2020 Decadal Report, each aperture design is off-axis with an inscribed primary mirror diameter of $\sim$6~m. The gaps between segments are fixed across the designs to 60~mm. The key aperture design parameters are listed in Table \ref{table:params}. Each aperture is matched to a circular unobscured Lyot stop, with an outer diameter (OD) set to 98$\%$ of the circle inscribed in the aperture perimeter. This design decision was informed by previous experience that APLC performance is best with a slightly undersized Lyot stop\cite{zimmerman16}.
\begin{table}[h!]
\centering
\begin{tabular}{ l l l l l l } 
\hline
\textbf{Design} & \textbf{1-Hex} & \textbf{2-Hex} & \textbf{3-Hex} & \textbf{4-Hex} & \textbf{5-Hex} \\
\hline
Circumscribed diameter (m) & 7.9445 & 7.2617 & 7.7231 & 7.1522 & 6.8526 \\
Inscribed diameter (m) & 6.0023 & 5.9994 & 5.9899 & 5.9937 & 5.9941 \\
Segment size (flat-to-flat) (m) & 2.6442 & 1.4475 & 1.1932 & 0.8511 & 0.6607 \\
Number of segments & 7 & 19 & 31 & 55 & 85\\
Gap size (m) & 0.006 & 0.006 & 0.006 & 0.006 & 0.006 \\
\hline
\\
\end{tabular}
\caption{Telescope aperture specifications for the 1-Hex, 2-Hex, 3-Hex, 4-Hex and 5-Hex designs.}
\label{table:params}
\end{table}
   
\begin{figure} [H]
   \begin{center}
   \begin{tabular}{c} 
   \includegraphics[height=7cm]{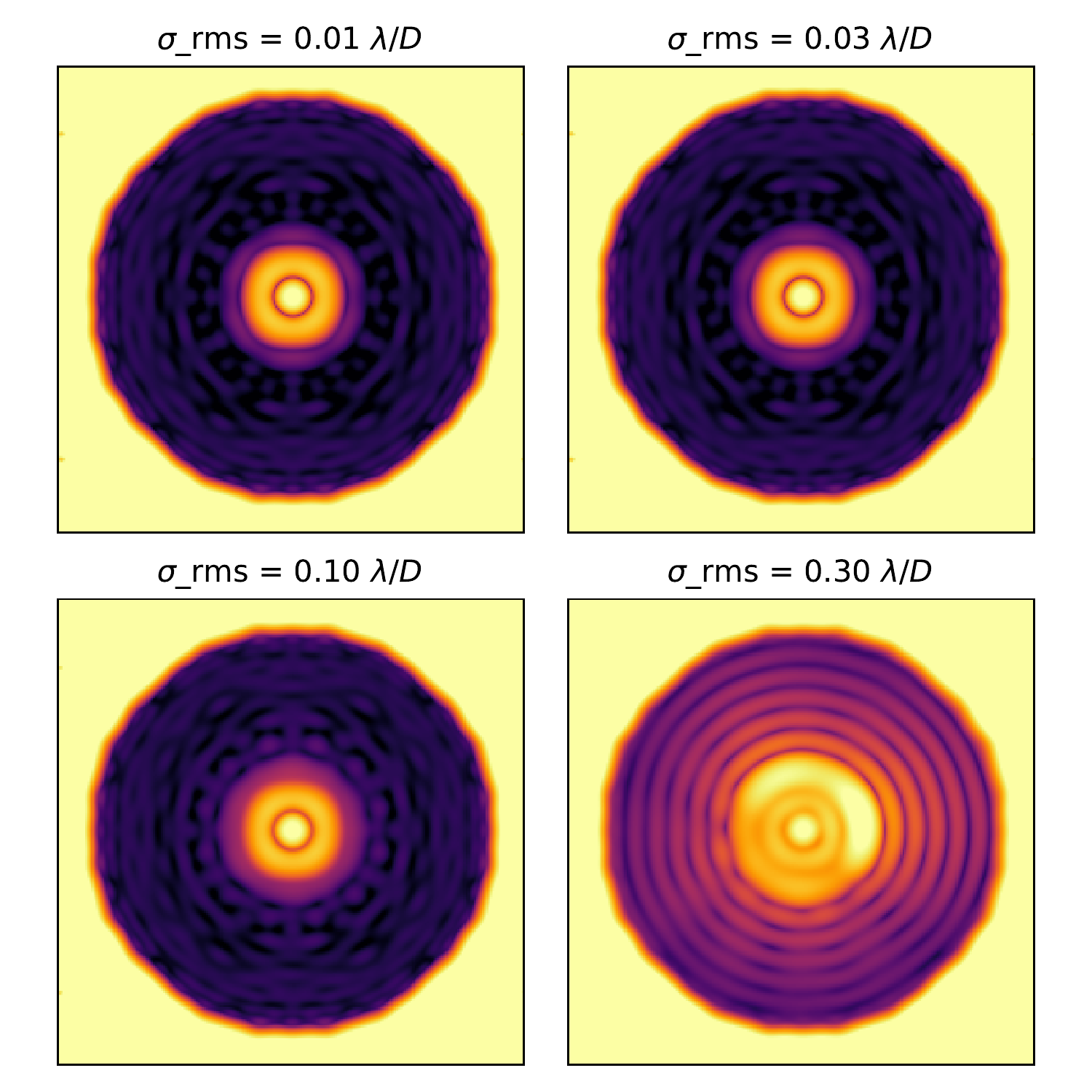}
   \includegraphics[height=6.7cm]{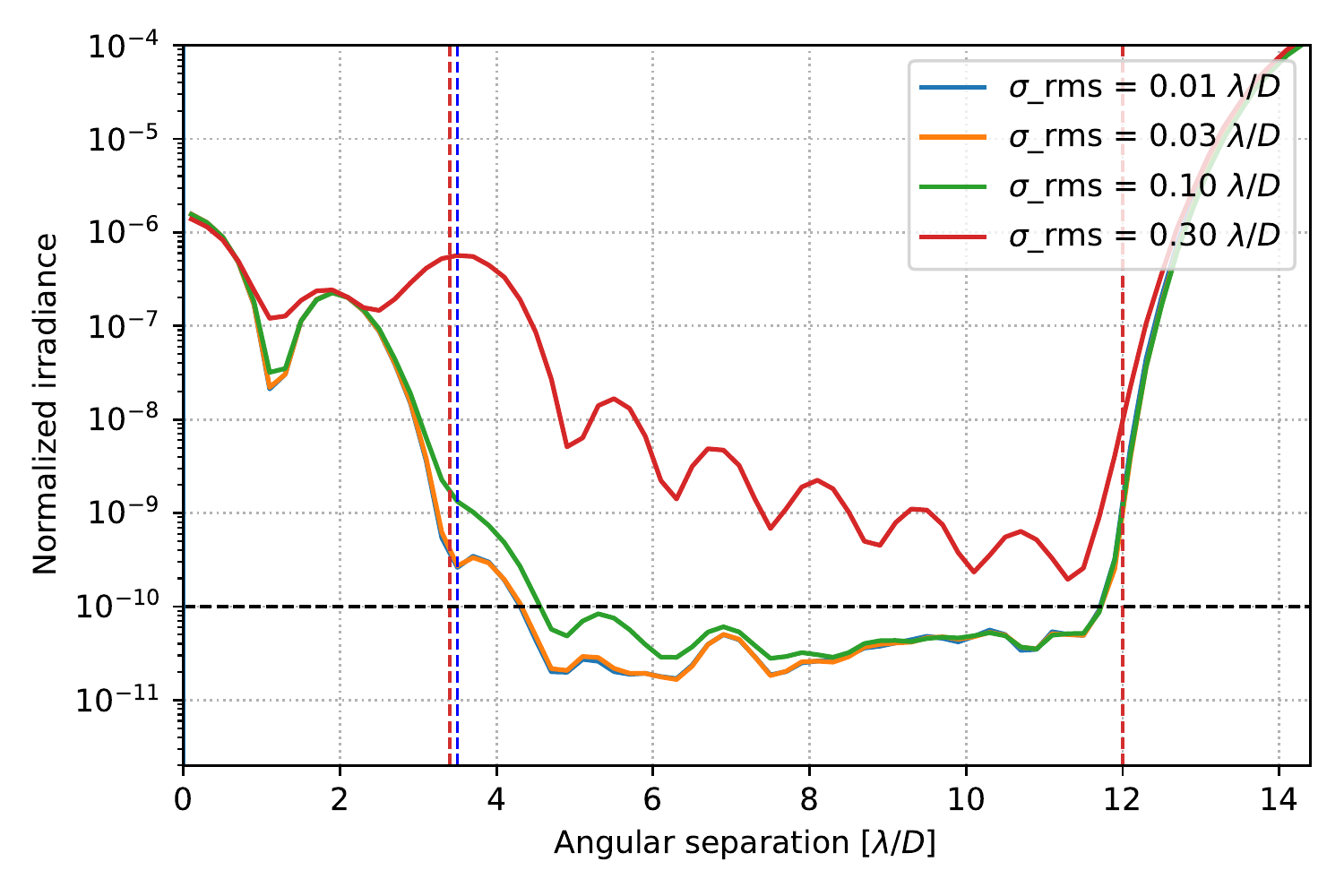}
   \end{tabular}
   \end{center}
   \caption[example] 
   { \label{fig:jitter} \textit{Left}: Broadband post-coronagraphic PSF images for the 5-Hex non-robust design for four representative levels of tip/tilt jitter. All images are shown on a logarithmic scale between $10^{-6}$ to $10^{-11}$. The blue and red vertical dashed lines delimit the FPM and bounds (IWA and OWA) on the dark zone, respectively. The black horizontal dashed line delimits the contrast goal. \textit{Right:} Normalized irradiance profiles (log scale) for the four levels of residual pointing jitter, averaged 100 wavelengths spanning the 10\% bandpass. We assume an isotropic distribution for tip-tilt offset. The different tip-tilt rms values were chosen to show the transition from no impact to significant impact on the normalized irradiance.}
   \end{figure}

The designs are optimized for a $10^{-10}$ contrast dark zone for a 10$\%$ relative spectral bandwidth. The outer perimeter of the annular dark zone (the effective OWA) is 12 $\lambda$/D. We set the inner edge of the dark hole to be smaller than the edge of the focal-plane mask in order to improve the design robustness to stellar diameter and low-order wavefront aberrations\cite{N'Diaye2015ApodizedPupilLyot}. Figure \ref{fig:jitter} illustrates the influence of tip-tilt jitter on the contrast performance for the 5-Hex non-robust design across the 10\% bandwidth, showing the design is robust to lower levels of tip-tilt jitter (achieving an average $10^{-10}$ contrast for a tip-tilt rms of $\sigma < 0.1 \lambda_{0}/D$), while for larger tip-tilt errors, the contrast in the dark zone starts to degrade. 

\begin{figure} [H]
   \begin{center}
   \begin{tabular}{c} 
   \includegraphics[width=\textwidth]{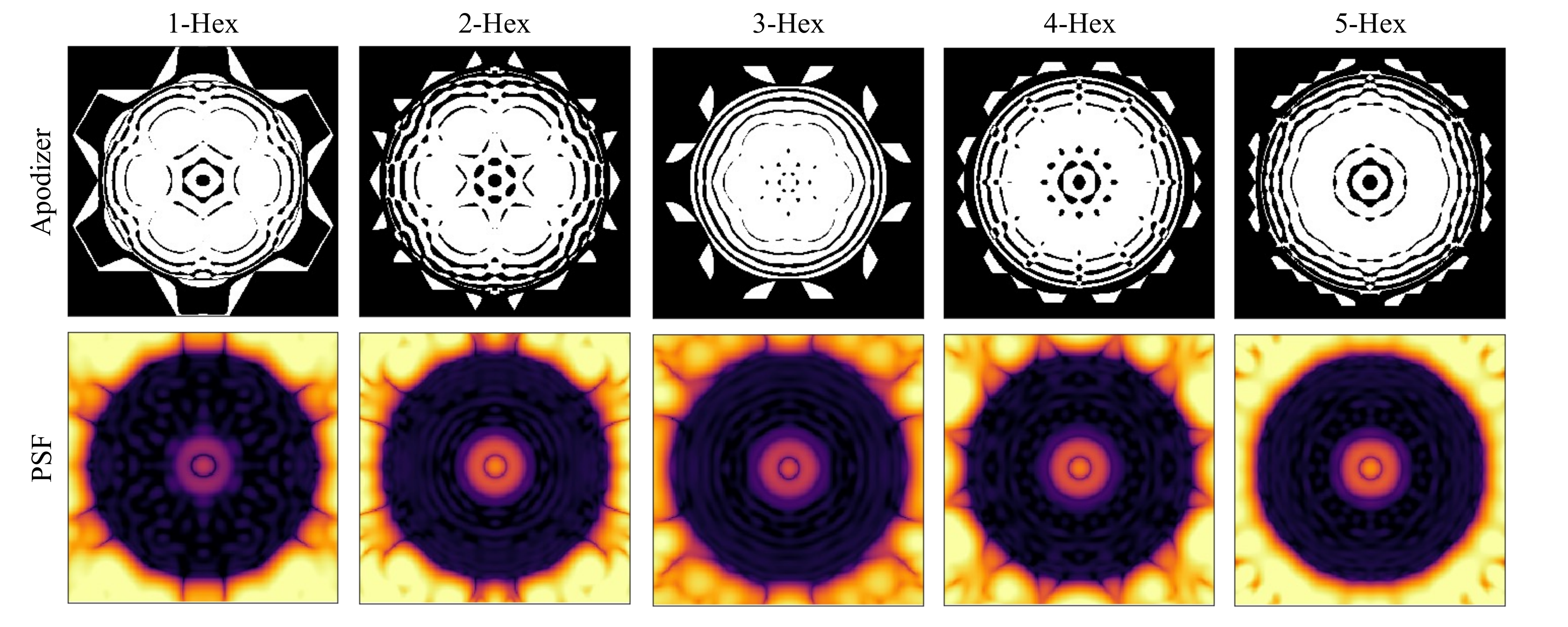}
   \end{tabular}
   \end{center}
   \caption[example] 
   { \label{fig:allhex}  \textit{Top}: The apodizer mask solutions (without built-in robustness properties) for each of the reference apertures shown in Figure~\ref{fig:hex_aps}. They are optimized for a 3.5 $\lambda$/D radius FPM and produce a $10^{-10}$ contrast in each pixel of the dark zone ranging between 3.4 and 12 $\lambda$/D in the coronagraphic image in 10\% broadband light. These masks are optimized to the manufacturing resolution of 1024 x 1024 pixels.
 \textit{Bottom}: The normalized irradiance of the post-coronagraphic stellar PSFs (on a logarithmic scale) for the non-robust apodizer mask solutions. The coronagraphic images for the non-robust apodizer masks for a 10\% spectral bandwidth. All are shown on logarithmic scale between 10$^{-4}$ and 10$^{-12}$.
 \newline}
   \end{figure}

\begin{figure} [H]
   \begin{center}
   \begin{tabular}{c} 
   \includegraphics[width=0.7\linewidth]{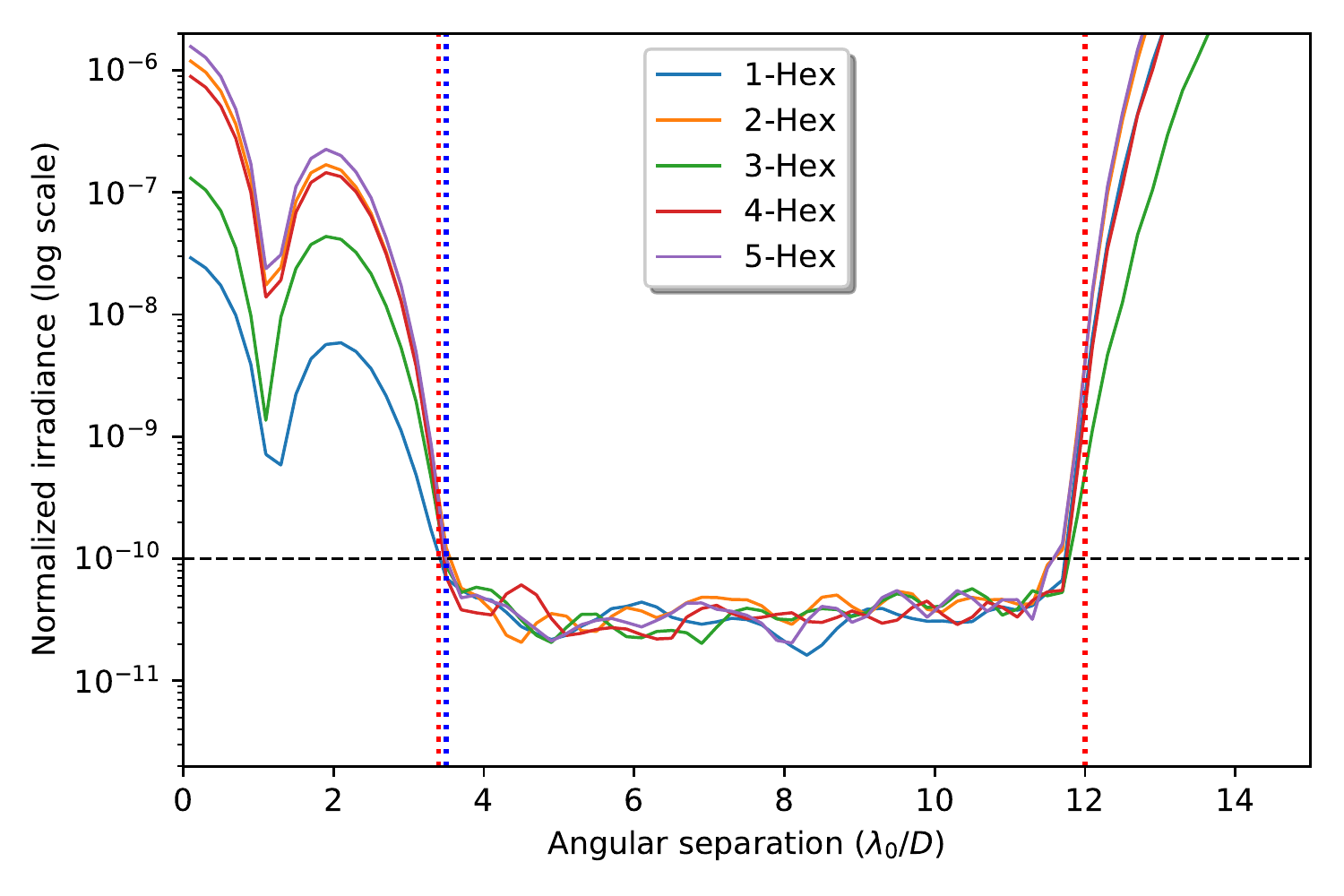}
   \end{tabular}
   \end{center}
   \caption[example] 
   { \label{fig:contrast} Normalized irradiance curves for the 1-Hex, 2-Hex, 3-Hex, 4-Hex, and 5-Hex non-robust designs, averaged over 3 wavelength samples spanning the 10\% bandpass. The radius of the focal-plane mask (set to 3.5$\lambda_{0}/D$) is delimited by the blue vertical dashed line, while the bounds (IWA and OWA, set to 3.4 and 12 $\lambda_{0}/D$) on the dark-zone region are delimited by the vertical red dashed line. The averaged contrast over the 10\% spectral bandpass in the dark zone region is below $10^{-10}$.}
   \end{figure}

The optimizations were run with these constraints at three wavelengths centered around $\lambda_0$ over a 10\% bandpass (note that while only three wavelengths are used for the optimizations, the broadband profiles illustrated in Figure~\ref{fig:allhex} are averaged over 11 wavelengths). Figure \ref{fig:allhex} shows both the apodizer mask solutions and the post-coronagraphic PSF images for the APLC designs discussed above. Figure \ref{fig:contrast} shows the radially averaged PSF profiles showing that all designs meet the 10$^{-10}$ contrast goal and exceed it at most angular separations.

\begin{figure} [H]
\begin{center}
\begin{tabular}{c} 
\includegraphics[width=0.7\linewidth]{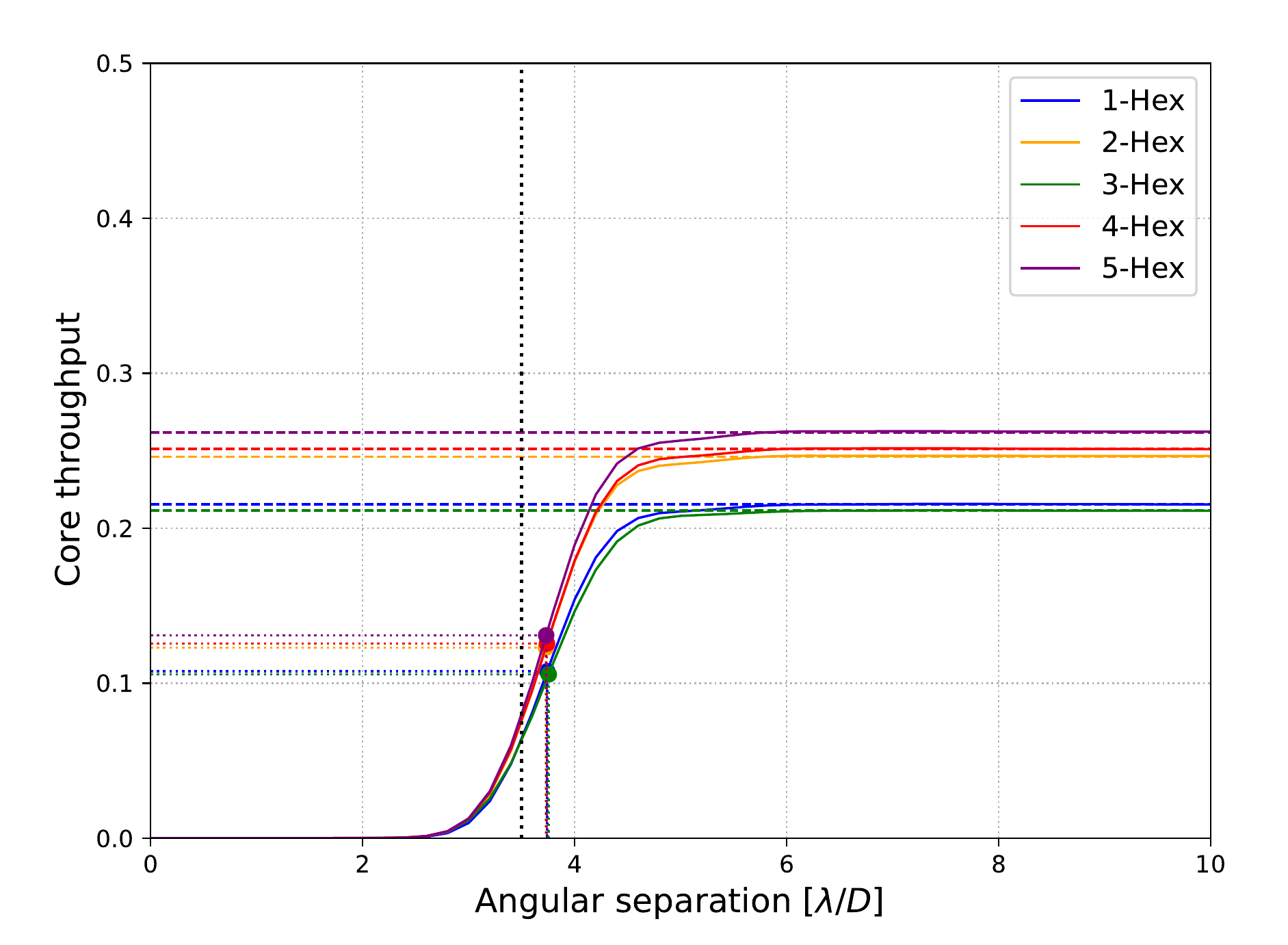}
\end{tabular}
\end{center}
\caption[example] 
{ \label{fig:throughput_all} Core throughputs for the non-robust APLC designs at 10\% bandpass, for each \textit{N}-Hex aperture shown in Figure~\ref{fig:hex_aps}. The horizontal dashed and dotted lines delimit the maximum and half-maximum core throughput for each design, respectively. The black vertical dashed line delimits the edge of the FPM, located at 3.5 $\lambda_0/D$. The 5-Hex design has the best throughput performance, where the aperture is the most circularized. Meanwhile the 1-Hex and 3-Hex designs exhibit the most throughput attenuation, where the ratio between the circumscribed and inscribed pupil diameter is much larger.}
\end{figure}

We show the core throughput for each of these designs in Figure~\ref{fig:throughput_all}, which is defined here as the ratio between encircled energies of the non-coronagraphic PSF and the off-axis (planet) coronagraphic PSF. We note the throughput performance between the 1-Hex and 3-Hex, and the 2-Hex and 4-Hex are very similar: both within 2\% of each other. This is because the design throughput is sensitive to the shape of the aperture perimeter. The 1-Hex and 3-Hex designs demonstrate the worst throughput performance due to their larger overall aperture diameter with respect to the diameter of the inscribed circle (ratio of 1.32 and 1.29, respectively). Meanwhile, the 5-Hex design, which has the smallest ratio between the inscribed and circumscribed diameter (1.14), has the best throughput performance. This is because the apodized pupil is directly related to the inscribed circle\cite{Kathryn19}. Light that is incident on the outer edges of the segmented primary, outside the inscribed diameter, is mostly discarded, resulting in a lower core throughput. 



\begin{figure} [htb!]
   \begin{center}
   \begin{tabular}{c} 
   \includegraphics[width=0.9\textwidth]{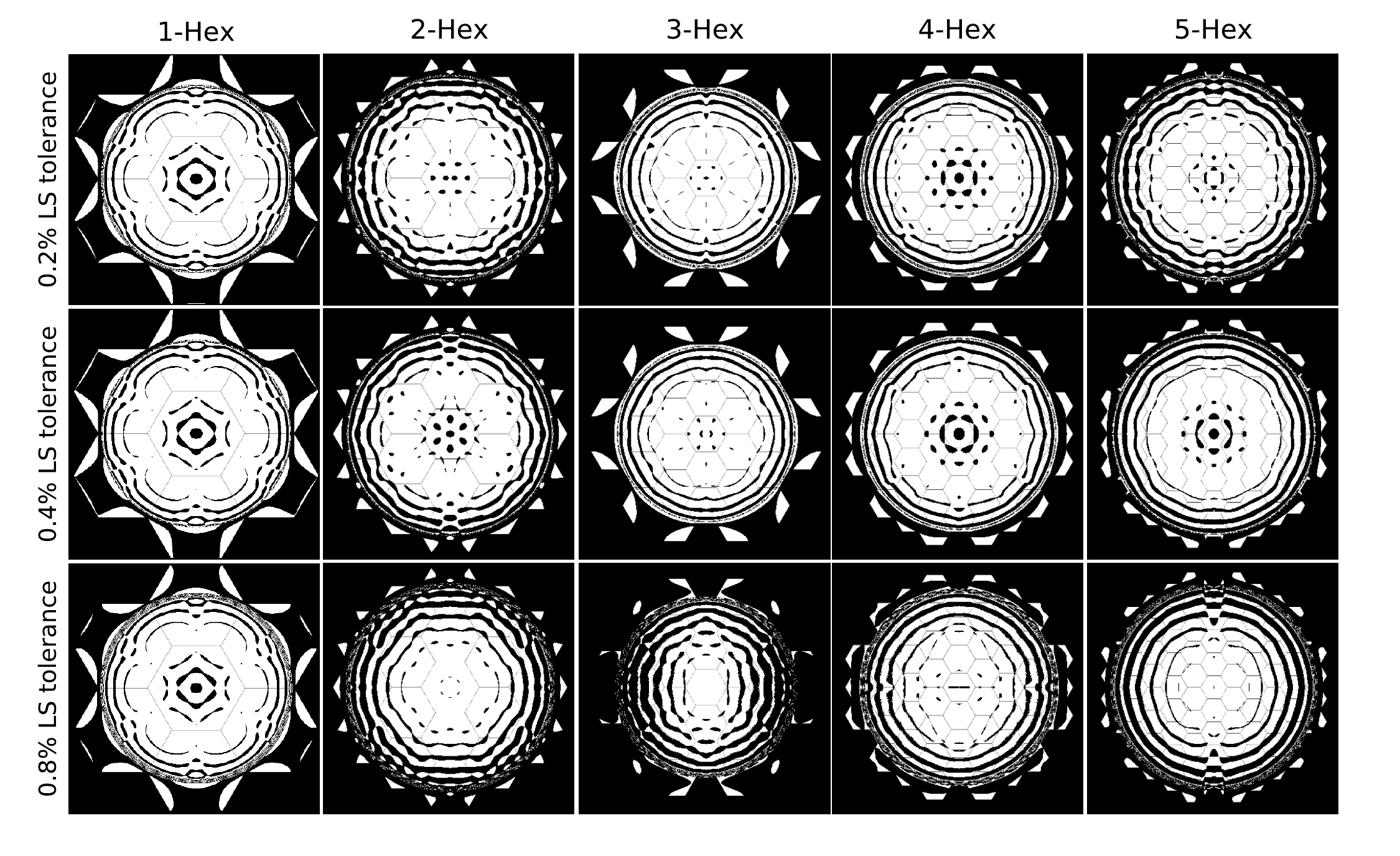}
   \end{tabular}
   \end{center}
   \caption[example] 
   { \label{fig:rob_apods} APLC apodizers produced by the N-Hex survey for Lyot stop misalignment tolerances of $\pm$0.2\% (\textit{top}), $\pm$0.4\% (\textit{middle}) and $\pm$0.8\% (\textit{bottom}). These designs produce annular, $10^{-10}$ contrast dark zones with working angle range 3.4 - 12 $\lambda_{0}/D$ over a 10\% bandpass. \textit{From left:} 1-Hex, 2-Hex, 3-Hex, 4-Hex, and 5-Hex.}
   \end{figure}

In addition to the ``non-robust'' designs above, we seek three further solutions for each of the apertures in Figure~\ref{fig:allhex}, optimized with 0.2$\%$, 0.4$\%$ and 0.8$\%$ Lyot stop misalignment tolerances. Figure \ref{fig:rob_apods} shows the apodizer solutions for each of the N-Hex telescope apertures with built-in robustness properties.  Differences in the apodization pattern are apparent where the optimizer has added features to accommodate Lyot stop translations in this range.

\begin{figure} [H]
   \begin{center}
   \begin{tabular}{c} 
   \includegraphics[width=0.7\textwidth]{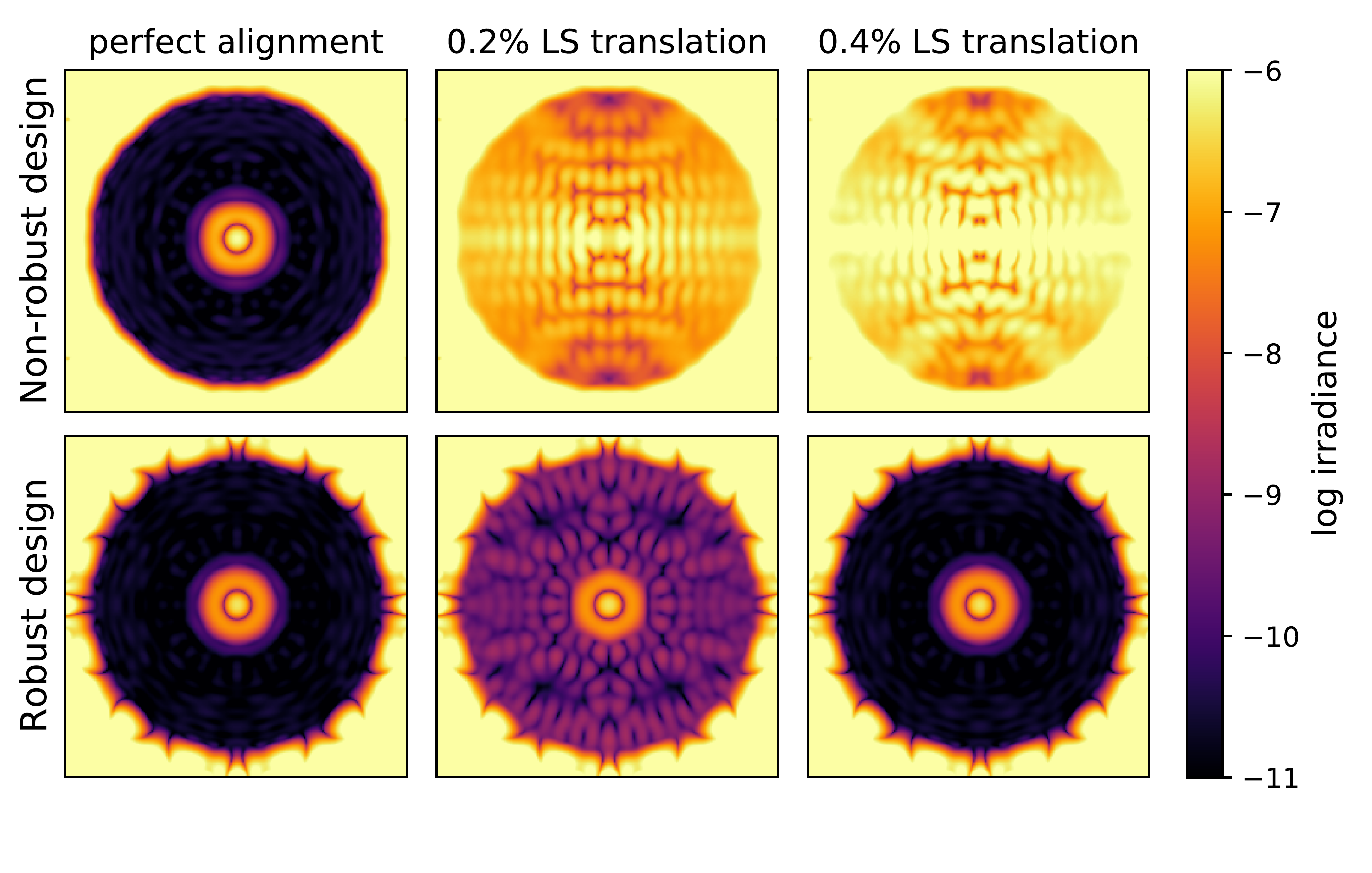}
   \end{tabular}
   \end{center}
   \caption[example] 
   { \label{fig:ls_robustness2} Post-coronagraphic PSF images for two 5-Hex APLC designs, one without built-in robustness (\textit{top}) and one optimized for robustness to Lyot stop translations of $\pm$0.4$\%$ (\textit{bottom}), shown first with perfect Lyot stop alignment, then with horizontal translations of 0.2\% and 0.4\% the pupil diameter applied to the Lyot stop. Clearly, the robust design is more capable of handling these small perturbations.  Note that while a $10^{-10}$ dark hole is not achieved at a 0.2\% LS translation in this design, in general, we can optimize a design for robustness at intermediate LS translations by including additional Lyot stops and positions in our LS optimization grid. This, however, comes at the cost of reduced throughput (see Figure~\ref{fig:throughput}).}
   \end{figure}

\begin{figure} [h]
   \begin{center}
   \begin{tabular}{c} 
   \includegraphics[width=\textwidth]{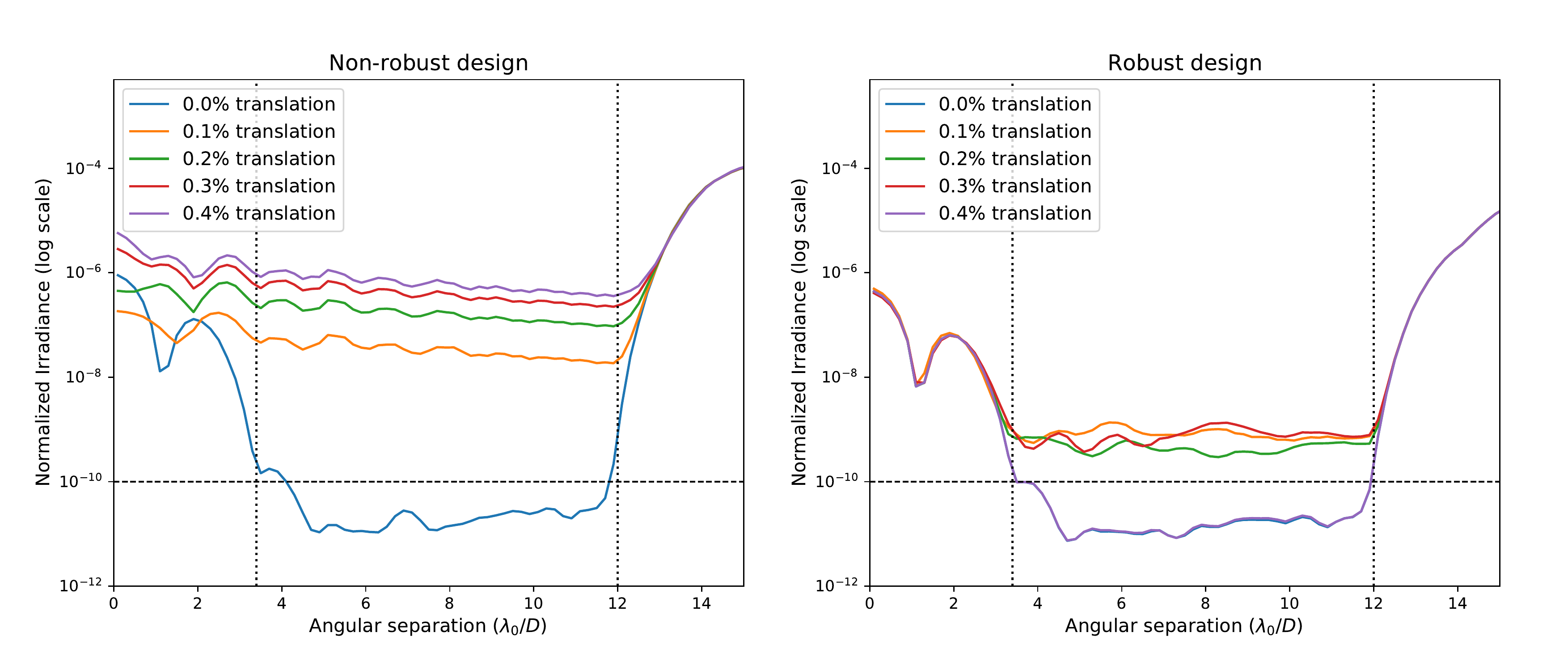}
   \end{tabular}
   \end{center}
   \caption[example] 
   { \label{fig:ls_robustness} Normalized irradiance profiles for two 5-Hex designs, one without (\textit{left}) and one with  robustness to Lyot stop misalignments of $\pm$0.4\% (\textit{right}), for x-axis Lyot stop translations from 0 to 0.4\% of Lyot stop size, with a step size of 0.1\%. The Normalized irradiance is averaged over 11 wavelengths spanning the 10\% bandpass. While the non-robust design produces a $10^{−10}$ contrast dark zone only for the case of 0\% translation of the Lyot stop, the robust design produces a 10$^{−10}$ contrast dark hole for different levels of Lyot stop misalignment.}
   \end{figure}

\begin{figure} [htb!]
   \begin{center}
   \begin{tabular}{c} 
   \includegraphics[width=0.7\linewidth]{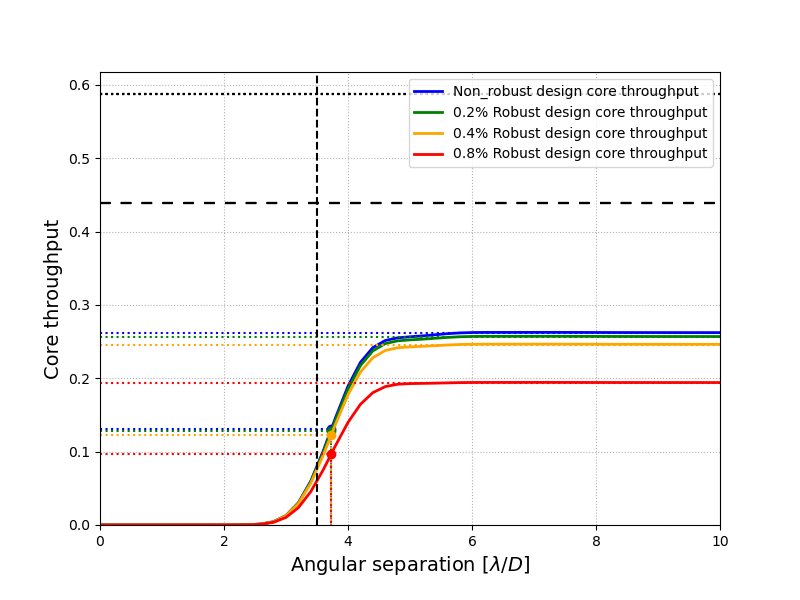}
   \end{tabular}
   \end{center}
   \caption[example] 
   { \label{fig:throughput} Core throughputs for four 5-Hex designs, one without Lyot stop robustness and three with $\pm$0.2\%, $\pm$0.4\% and $\pm$0.8\% Lyot stop alignment tolerances, respectively. The plots illustrate the trade-off between design robustness and throughput - increased design robustness results in a corresponding decrease in throughput. Indeed, there is a throughput reduction of 26\% between the non-robust design and the design optimized to a LS misalignment tolerance of 0.8\%.}
   \end{figure}

 Compared to the non-robust designs, which are optimized for a single on-axis Lyot stop, the robust designs are optimized for a set of 9 Lyot stops in a three-by-three grid centered around the on-axis position. Figure~\ref{fig:ls_robustness2} and \ref{fig:ls_robustness} show examples of the post-coronagraphic PSF images and normalized irradiance profiles for two designs (5-Hex geometry), one with and one without built-in Lyot stop robustness properties, in the presence of different horizontal translations of the Lyot stop. They illustrate that while the non-robust design generates a dark zone only for a perfectly centered Lyot stop, the robust design produces dark holes for different Lyot stop offsets, illustrating its ability to yield contrast in the presence of a slightly decentered Lyot stop. While these solutions are very promising in terms of robustness to LS misalignments, they come at the cost of decreased throughput. Figure~\ref{fig:throughput} shows the Core throughput for the four 5-Hex APLC designs, illustrating the trade-off between design robustness and core throughput. 
 

\section{Comparative Wavefront Error Sensitivity Analysis}
\label{sec:pastis} 
In our previous analysis of the N-Hex designs, we assumed an idealized scenario of a system without any wavefront errors. In a real optical system however, aberrations will lead to a degradation of image quality. Coronagraphic instruments are particularly sensitive to wavefront errors in the optical system, which contaminate the focal-plane image. For segmented telescopes, a significant component for wavefront errors is misalignment between individual segments. In this section, we study the performance of the five N-Hex aperture and coronagraph designs to these kinds of segment-level aberrations. In particular, we use the PASTIS sensitivity analysis to set segment-level piston tolerances\cite{2021JATIS...7a5004L, 2019zndo...3382986L}. We then expand this analysis to custom thermal aberration modes on the segments, which provides the tolerances for thermal gradients for a requested static target contrast.

\begin{figure} [ht]
   \begin{center}
   \includegraphics[width=\textwidth]{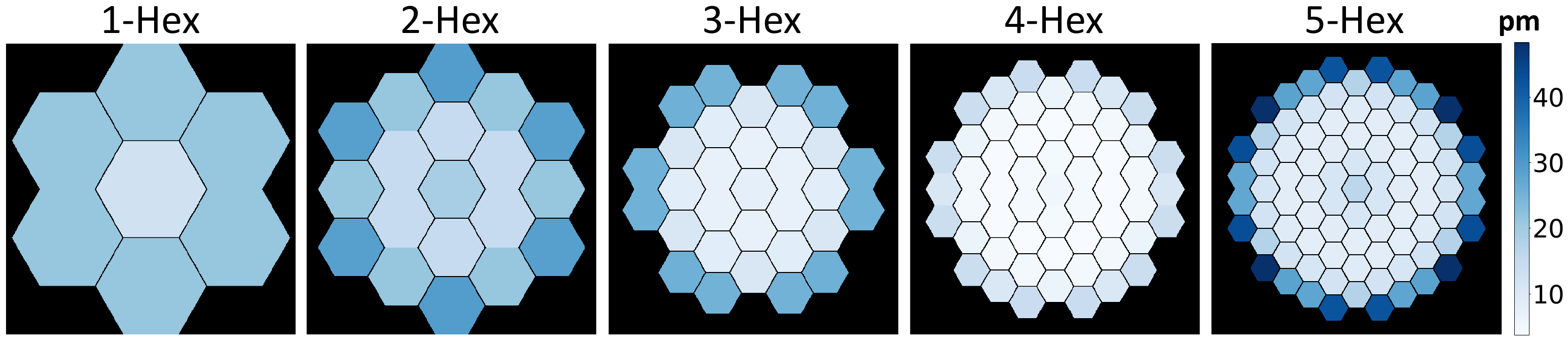}
   \end{center}
   \caption[example] 
   { \label{fig:piston_tolerances_pastis} Static piston tolerance for each design from 1-hex to 5-hex, which derives from the PASTIS sensitivity analysis\cite{2022A&A...658A..84L,2021JATIS...7a5004L,2019zndo...3382986L}.  The values for each of these segment correspond to the standard deviation of the surface piston error for each segment with a target contrast of $10^{-10}$. All five tolerance maps are plotted on the same scale.}
   \end{figure}

The PASTIS approach includes building a calibration matrix by propagating known input wavefront aberrations through an end-to-end model of the telescope and coronagraph, and measuring the dark-hole intensity. We then inverse this calibration matrix to calculate the tolerances required for each segment, expressed as a standard deviation. We produce spatially dependent requirement maps (instead of describing the wavefront tolerance globally for the total pupil). Figure \ref{fig:piston_tolerances_pastis} shows the segment-level piston tolerances for the five designs required to achieve a statistical mean contrast of $10^{-10}$ in the coronagraphic dark zone, over a normally distributed set of aberration maps.

\begin{figure} [ht]
   \begin{center}
   \includegraphics[width=0.5\textwidth]{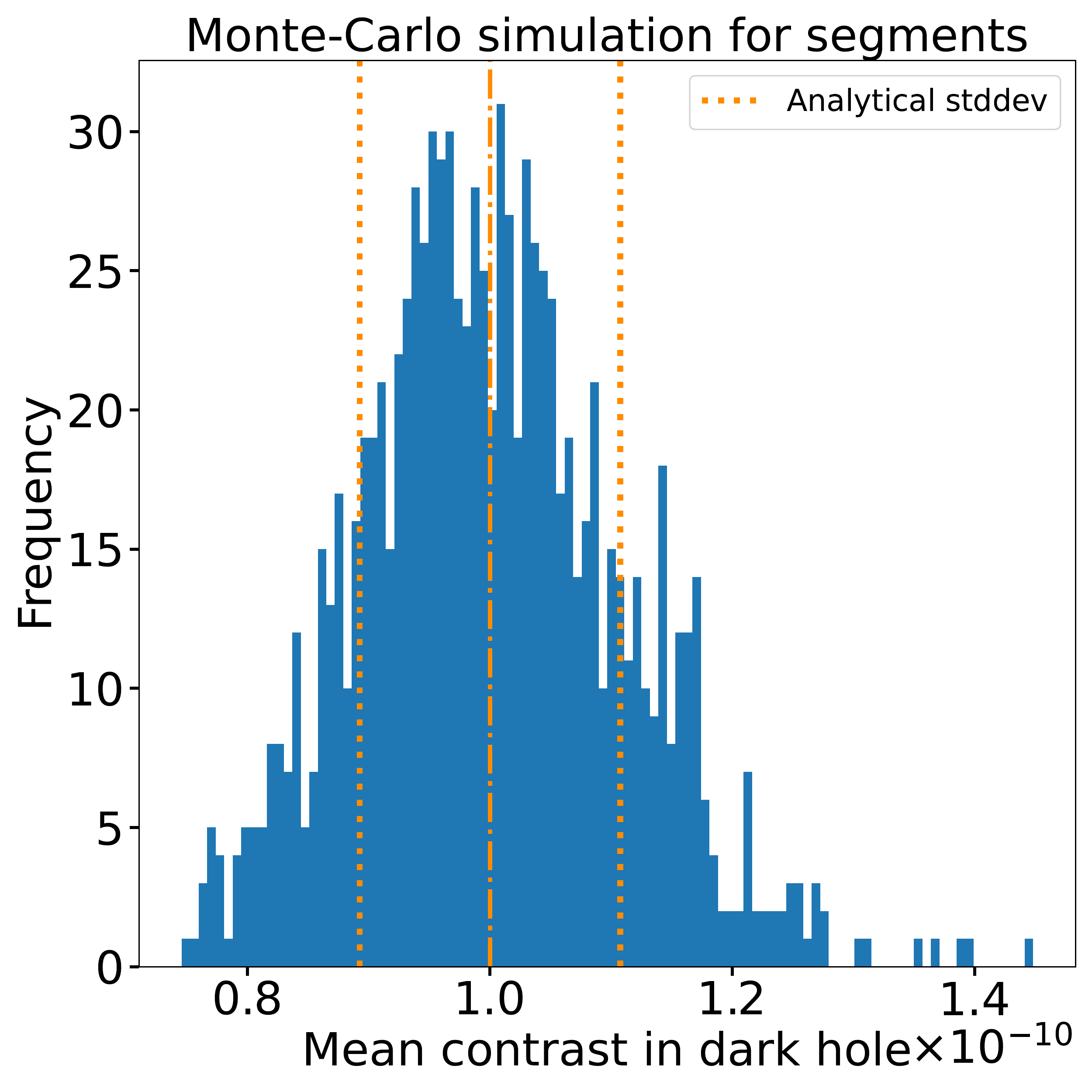}
   \end{center}
   \caption[example] 
   { \label{fig:Monte_carlo_Pastis_piston} Validation of the segment-level piston requirements for one example (5-Hex) from Figure\ref{fig:piston_tolerances_pastis}. In this simulation, random piston maps are generated with zero-mean normal distributions and to the per-segment standard deviations from the tolerance map in Fig.\ref{fig:piston_tolerances_pastis}. These are then propagated numerically through an optical simulator\cite{2022A&A...658A..84L,2021JATIS...7a5004L} and the resulting average DH contrast is recorded as one data point. The distribution shown here corresponds to a sample size of $n=1000$ such measurements. Its mean value and standard deviation match both the expected mean of $10^{-10}$ contrast in the dark zone (dashed-dotted line), as well as the analytically predicted standard deviation (dashed lines).  All five designs produce very similar histograms not shown here.}
   \end{figure}

Out of the five designs considered, we find that the 4-Hex design has the most stringent requirements, and the 5-Hex has the most relaxed piston requirements overall. In general, we find that for the designs with increased segment numbers, the inner segments have more stringent tolerance requirements than the outer ones.  We also observe that the spread between the most and least tolerant segments is the largest for the 5-Hex design, while it is the smallest for the 4-hex design. In a next step, we validate these tolerance maps by performing end-to-end optical Monte Carlo (MC) simulations. For the validation of one tolerance map, we draw $n=1000$ segmented aberration maps, where the coefficient of the piston aberration is drawn randomly from a normal distribution with a zero mean and standard deviation given by the respective map. These aberration realizations are then propagated through an optical simulator and their average DH contrast measured. The distribution of these measurements is shown in the histogram in Figure~\ref{fig:Monte_carlo_Pastis_piston}.
In the resulting histogram, we calculate a mean value that matches the requested statistical mean contrast, and the standard deviation of the contrast distribution is equal to the analytically calculated value of this optical system as computed with the PASTIS model\cite{2021JATIS...7a5004L}.

\begin{figure} [htb!]
   \begin{center}
   \begin{tabular}{c}  
   \includegraphics[width=\linewidth]{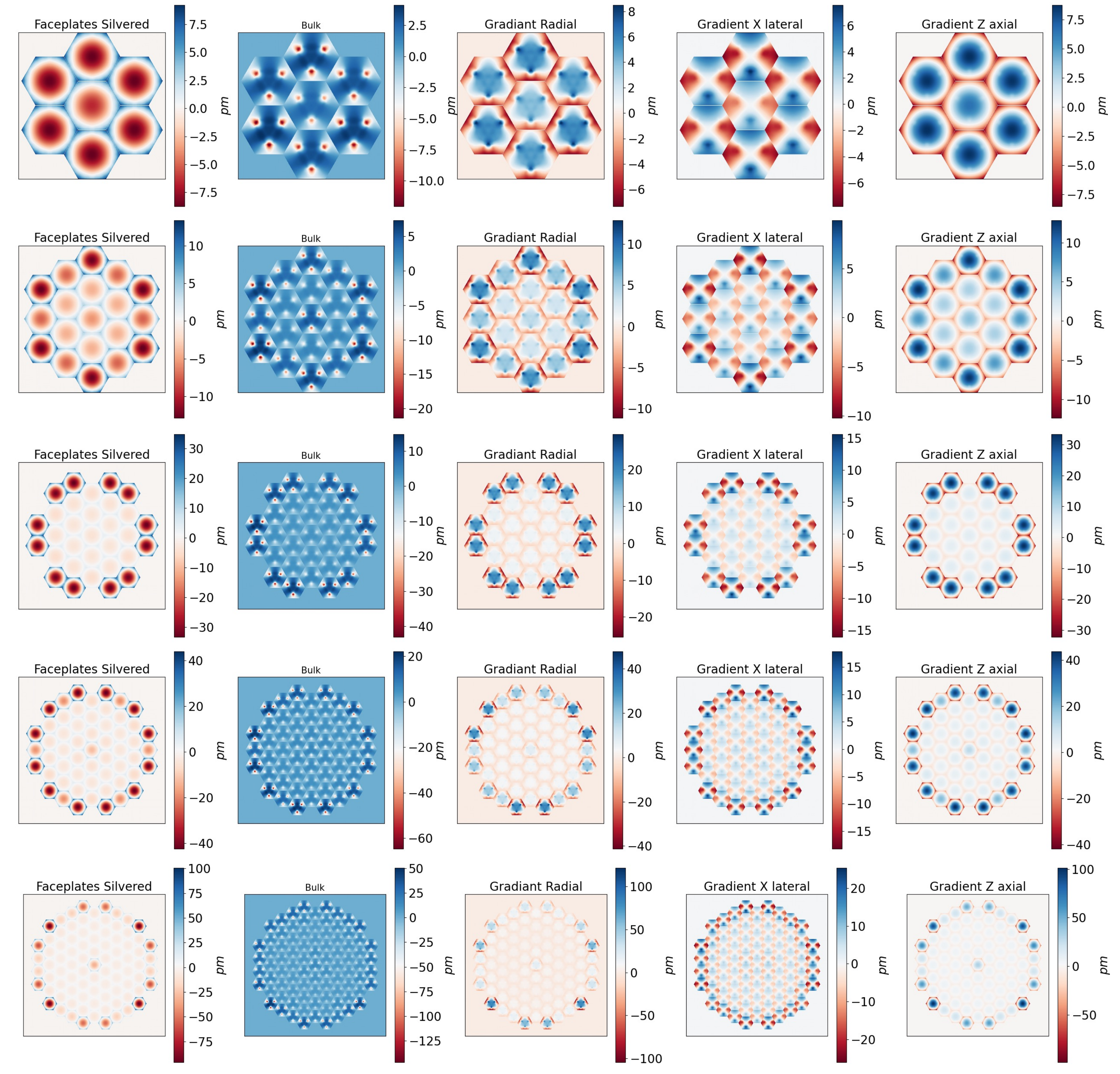}
   \end{tabular}
   \end{center}
   \caption[example] 
   { \label{fig:Harris_surface_maps_all-Hex} Static surface tolerances caused by temperature changes on the individual segments for five different hexagonal aperture geometries and APLC designs. These results are based on the PASTIS sensitivity analysis of all five thermal modes combined (faceplate, bulk, and thermal gradients)\cite{sahoo2022}, and we obtain the temperature tolerance for each segment to reach $10^{-10}$ contrast in the dark hole. }
   \end{figure}

We extend the PASTIS approach to study the thermo-elastic effect associated with the primary mirror back plane support structure under thermal influence. The mirror mounting pads have direct contact with the mirror substrate and induce surface deformations on the order of picometers which directly impacts the performance of the coronagraph. We use five kinds of segment-level surface disturbances obtained from L3 Harris Technologies\cite{sahoo2022} as the finite elements for our analysis. These finite elements are the segment-level response when a 1~mK temperature gradient is applied along different axial directions of a segment; they are more likely to occur than segment-level Zernike polynomials. We use these finite elements in the PASTIS sensitivity analysis like in the previous paragraph and derive segment-level temperature requirements to maintain an average dark hole contrast of $10^{-10}$. The results in terms of surface deformation errors are shown in Figure~\ref{fig:Harris_surface_maps_all-Hex}.

\begin{figure} [hb]
   \begin{center}
   \begin{tabular}{c} 
   \includegraphics[width=\linewidth]{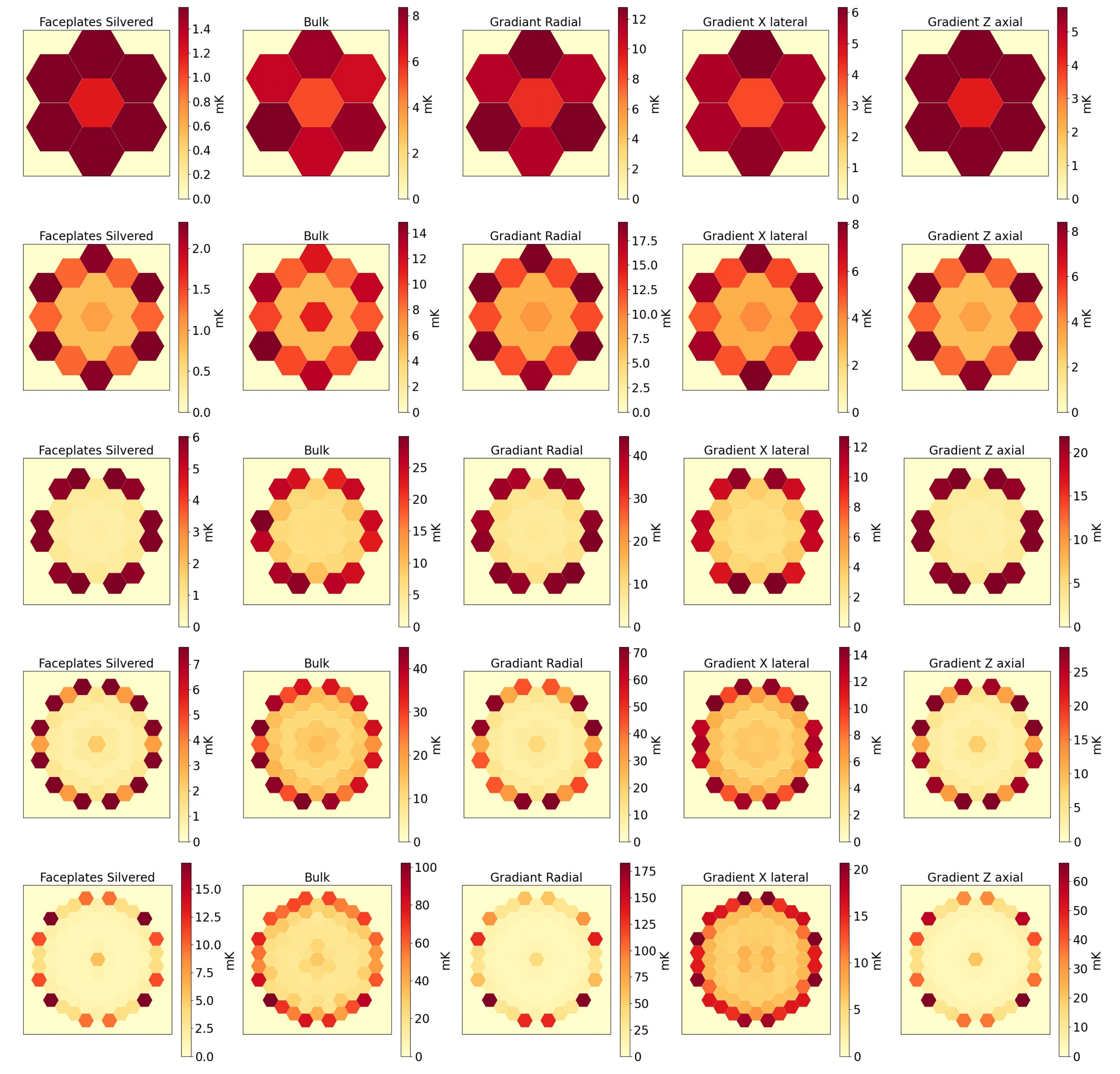}
   \end{tabular}
   \end{center}
   \caption[example] 
   { \label{fig:Harris_Themal_maps_all-Hex} Static, per-segment temperature gradient tolerances for five different hexagonal aperture geometries and APLC designs. These results are based  on the PASTIS mode analysis of all five thermal modes combined (faceplate, bulk, and thermal gradients) for currently envisioned hexagonal segments\cite{sahoo2022}, and we obtain the temperature tolerance for each segment to reach $10^{-10}$ contrast in the dark hole. }
   \end{figure}

Since the thermal aberration modes establish a relation between the influence of a thermal gradient and the resulting surface deformation, we can use them to convert the surface requirements in Fig.~\ref{fig:Harris_Themal_maps_all-Hex} into requirements expressed as a tolerable temperature change standard deviation. These physical-unit tolerance maps are displayed in Figure~\ref{fig:Harris_Themal_maps_all-Hex} for all five designs.



\section{Conclusions}
\label{sec:conclusion}
In this paper, we have presented APLC-Optimization, a coronagraph design survey toolkit written in Python for optimizing and exploring APLC solutions for any telescope geometry and science goal. This toolkit simplifies the organization, execution and analysis of extensive design parameter space surveys. It allowed us to establish relationships and trade-offs between design parameter combinations and identify new design approaches for telescopes with various performance requirements. A notable feature of the design toolkit is the possibility to include robustness to LS misalignments as an optimization parameter. In this way, the resulting apodizers still perform reasonably well in terms of contrast for a small range of off-center LS positions.

We showed examples of apodizer masks optimized by this toolkit for HiCAT and Gemini/GPI, illustrating its ability to produce competitive designs for both space- and ground-based high-contrast imaging instruments. This includes the capability to handle segmentation and secondary support features with excellent contrast performance and optimized planet throughput in 10-15\% bandwidths.

Finally, we presented an application of the toolkit for the case of a 6~m aperture off-axis segmented telescope. We performed a design survey investigating the effect of varying segment sizes on the overall coronagraph performance. The investigated telescope designs all have a 6~m inscribed diameter aperture with a varying number of hexagonally segmented rings, from $N=1$ to $N=5$. Each of these five apertures were used to create optimized apodizers with and without baked-in robustness to LS misalignments.

While all of the prescribed APLC solutions reach the design DH contrast of $10^{-10}$ and a reasonable throughput, we evaluated the five designs through a WFE sensitivity analysis used previously to derive the per-segment WFE tolerances of LUVOIR-A. In a detailed study for piston-only local aberrations, the PASTIS sensitivity analysis let us derive an individual WFE standard deviation per segment. Following the extended tolerancing approach of this method presented in Ref.~\citenum{sahoo2022}, we derived first physical tolerances for thermal gradients.

These preliminary results indicate promising performance for the APLC on the next large space-based IR/O/UV flagship. The comparative analysis between the designs indicates that the 5-Hex design is the most promising in terms of maximum core throughput and WFE tolerance; however, further work is necessary to present a more thorough examination. Applications like these underline the relevance of optimization tools like the one presented in this paper, as they will enable us to expand on these pre-fatory results in the near future.

For more quantitative details on the methodology, see Sahoo et al. 2022 \cite{sahoo2022}. 

\clearpage
\acknowledgments 
This work was funded in part by NASA Exoplanet Exploration Program’s (ExEPs) Segmented Coronagraph Design and Analysis (SCDA) study. E.H.P. is supported by the NASA Hubble Fellowship grant \#HST-HF2-51467.001-A awarded by the Space Telescope Science Institute, which is operated by the Association of Universities for Research in Astronomy, Incorporated, under NASA contract NAS5-26555. 
The HiCAT optimization work was supported in part by the National Aeronautics and Space Administration under Grant 80NSSC19K0120 issued through the Strategic Astrophysics Technology/Technology Demonstration for Exoplanet Missions Program (SAT-TDEM; PI: R. Soummer). 
The GPI2.0 optimization work was funded by the STScI Discretionary Research Fund (D0101.90238). 
I.L. acknowledges the support by a postdoctoral grant issued by the Centre National d'Études Spatiales (CNES) in France. The Gemini Planet Imager 2.0 project upgrade is supported by the National Science Foundation under Grant No. AST-1920180, and also significantly supported by the Heising-Simons Foundation. A.S. acknowledges the support by the Ultra-Stable Telescope Research and Analysis (ULTRA) Program under Contract No. 80MSFC20C0018 with the National Aeronautics and Space. 
This research made use of HCIPy, an open-source object-oriented framework written in Python for performing end-to-end simulations of high-contrast imaging instruments.

\bibliography{main} 
\bibliographystyle{spiebib} 

\end{document}